\documentclass{appolb}
\usepackage{epsfig}

\def\permil{\%\raise.10ex\hbox{$_{\scriptstyle 0}$}}

\def\beq{\begin{equation}}
\def\eeq{\end{equation}}
\def\beqn{\begin{eqnarray}}
\def\eeqn{\end{eqnarray}}

\newcommand{\ki}{{\bf k}_i}
\newcommand{\kim}{{\bf k}_{i-1}}

\newcommand{\kj}{{\bf k}_j}

\newcommand{\ka}{{\bf k}_a}

\newcommand{\kjet}{{\bf k}_J}

\newcommand{\del}[1]{\delta^{(2)}\left(#1\right)}

\newcommand{\qi}{{\bf q}_i}\newcommand{\qip}{{\bf q}_{i+1}}
\newcommand{\qj}{{\bf q}_j}

\newcommand{\qone}{{\bf q}_1}\newcommand{\qtwo}{{\bf q}_2}
\newcommand{\qa}{{\bf q}_a}\newcommand{\qb}{{\bf q}_b}

\begin{document}
\title{The High Energy Limit of QCD: BFKL Cross Sections%
\thanks{Presented at School on QCD, Low $x$ Physics, Saturation and Diffraction, Copanello (Calabria, Italy), July 1--14 2007}%
}
\author{Agust{\' \i}n Sabio Vera\address{Physics Department, Theory Division, CERN, CH-1211 Geneva 23, Switzerland}}
\maketitle
\begin{abstract}
In this contribution we describe in some detail different aspects of the 
construction of BFKL cross sections. We focus on several effects which are 
relevant at next--to--leading order. In particular, we describe QCD coherence 
in DIS final states, improvements of the collinear region in multi-Regge 
kinematics, inclusive jet production at next--to--leading order, and 
azimuthal angle decorrelations of Mueller--Navelet jets at hadron colliders. 
\end{abstract}
  
\section{Introduction}
When dealing with the description of scattering amplitudes at very high 
center--of--mass energies a very useful formalism is the 
Balitsky--Fadin--Kuraev--Lipatov (BFKL) 
approach~\cite{BFKL1,BFKL2,BFKL3,BFKL4,BFKL5}. In the Regge limit the dominant 
degrees of freedom are $t$--channel ``Reggeized'' gluons which interact with 
each other via standard gluons in the $s$--channel and a gauge invariant 
Reggeized gluon - Reggeized gluon - gluon vertex.
This picture emerges as a consequence of multi--Regge kinematics where 
gluon evolution takes place with ordering in longitudinal components 
but not in transverse momenta.  
At very high energies this structure should be modified to include 
unitarization corrections. However, there should be a window at present 
and future colliders where the BFKL predictions provide a good description 
of the experimental data.

When terms of the form $(\alpha_s \ln{s})^n$ are resumed we are in the 
leading--logarithmic approximation (LLA). In this limit the strong coupling 
does not run and ${\alpha}_s$ is a constant parameter. 
We should have written $\ln{s/s_0}$ but it turns out that in the LLA we are 
free to choose any $s_0$. This means that there is a lot of freedom in the 
LLA when trying to fit experimental data. This theory is much more 
constrained when we include terms of the form $\alpha_s (\alpha_s \ln{s})^n$. 
In this next--to--leading logarithmic approximation (NLLA) the coupling is 
allowed to run and the energy scale $s_0$ has to be determined. As a matter 
of fact, cross sections are constructed to be independent of $s_0$ at NLO. 
However, different choices of this scale do affect higher orders in the 
resummation which might be important when making BFKL predictions. 

In the coming sections we discuss several important aspects to take into 
account when constructing BFKL cross sections beyond the LLA. 
In section~\ref{coherence} we describe final states at small 
Bjorken $x$ in Deep Inelastic Scattering (DIS). The idea of colour coherence 
is introduced and its implementation in the CCFM equation discussed. 
The predictions for jet rates are the same at LO when the CCFM or BFKL 
equations are used and we explain why. In section~\ref{collinear} it is shown 
that the region of applicability of multi--Regge kinematics can be extended to also include regions with collinear emissions. In this case there exists an 
interesting structure at higher orders which can be cast into a Bessel function of the first kind accounting for double logarithms in transverse scales. 
Up to the NLLA this double logarithms 
also appear in the inclusive production of 
a jet centrality emitted in rapidity at a hadron collider. In this case 
impact factors and the emission vertex of the central jet have to modified. 
This is explained in some detail in section ~\ref{inclusivejet}.
Finally, in section~\ref{conformal} we show how the $SL(2,C)$ invariance 
present in the BFKL hamiltonian for non--zero momentum transfer appears in 
the azimuthal angle dependence of multijet events. As an example we discuss 
the case with two hard external scales of Mueller--Navelet jets at a hadron 
collider.

\section{DIS final states at small $x$ and the CCFM equation}
\label{coherence}

In QED coherence suppresses soft bremsstrahlung from electron--positron pairs. 
In QCD processes such as $g \rightarrow q {\bar q}$ soft gluons at an angle 
from one of the fermionic lines larger than the angle of emission in the 
$q{\bar q}$ pair resolve the total colour charge of the pair. This is the 
same as that of the parent gluon and radiation occurs as if the soft gluon 
was emitted from it. This ``colour coherence'' can be put as angular ordered 
sequential gluon emissions. 

If the $(i-1){\rm th}$ emitted gluon from the proton in DIS has energy 
$E_{i-1}$, then a gluon radiated from it with a fraction $(1-z_{i})$ of its 
energy and a transverse momentum $q_{i}$ has opening angle 
\begin{eqnarray}
\theta_{i}\approx \frac{q_{i}}{(1-z_{i})E_{i-1}}, \,\,\,\,
z_{i}=\frac{E_{i}}{E_{i-1}}.
\end{eqnarray}
Colour coherence leads to angular ordering with increasing opening angles 
towards the hard scale (the photon). We then have \(\theta_{i+1} 
> \theta_{i}\), or 
\begin{eqnarray}
\frac{q_{i+1}}{1-z_{i+1}}>\frac{z_{i}q_{i}}{1-z_{i}},
\end{eqnarray}
which reduces to $q_{i+1}>z_{i}q_{i}$ in the limit $z_{i},z_{i+1}\ll1$. 
In~\cite{Catani:1990yc,Marchesini:1995wr,Ciafaloni:1988ur,Catani:1991gu} 
the BFKL equation for the unintegrated structure function was obtained in a form suitable for the study 
of exclusive observables: 
\begin{eqnarray}
f_{\omega}(\mbox{\boldmath $k$})=f_{\omega}^{0}(\mbox{\boldmath $k$})
+\bar\alpha_{S}\int\frac{d^{2}\mbox{\boldmath $q$}}{\pi q^{2}}\int_{0}^{1}
\frac{dz}{z}z^{\omega}\Delta_{R}(z,k)\Theta(q-\mu)f_{\omega}(\mbox{\boldmath 
$q$}+\mbox{\boldmath $k$}).
\end{eqnarray}
$\mu$ is a collinear cutoff, \mbox{\boldmath $q$} the transverse 
momentum of the emission, and 
\begin{eqnarray}
\Delta_{R}(z_{i},k_{i})=\exp\left[-\bar\alpha_{S}\ln\frac{1}{z_{i}}\ln
\frac{k_{i}^{2}}{\mu^2}\right],
\end{eqnarray}
with \(k_i\equiv|\mbox{\boldmath $k$}_{i}|\), and $\bar\alpha_{S}\equiv \alpha_{S} N_c /\pi$. This expression predicts gluon emissions with the virtual 
corrections summed to all orders. Since 
$f_{\omega}$ is an inclusive structure function, it includes the sum over 
final states. After this sum the $\mu$-dependence cancels.

To get the structure function we integrate over $\mu^{2} \leq q_{i}^{2}\leq Q^{2}$:
\begin{eqnarray}
F_{0\omega}(Q,\mu) \equiv \Theta(Q-\mu) + \sum_{r=1}^{\infty}\int_{\mu^{2}}
^{Q^{2}}\prod_{i=1}^{r}\frac{d^{2}\mbox{\boldmath $q$}_{i}}{\pi q_{i}^{2}}
dz_{i}\frac{\bar{\alpha}_{S}}{z_{i}}z_{i}^{\omega}\Delta_{R}(z_{i},k_{i}),
\end{eqnarray}
with $i$ being each gluon emission. A fixed number $r$ of emitted gluons gives 
\begin{eqnarray}
F_{0\omega}(Q) = \int_{0}^{1} dx ~x^{\omega} F_{0}(x,Q) = 1 + 
\sum_{r=1}^{\infty} F_{0\omega}^{(r)}(Q).
\end{eqnarray}
The expansion for $F_{0\omega}^{(r)}(Q,\mu)$ reads~\cite{Marchesini:1995wr}
\begin{eqnarray}
F^{(r)}_{0\omega}(Q,\mu)=\sum_{n=r}^{\infty}C^{(r)}_{0}(n;T)
\frac{\bar\alpha_{S}^{n}}{\omega^{n}},
\end{eqnarray}
with $T \equiv \ln({Q/\mu})$. Therefore:
\begin{eqnarray}
F_{0\omega}(Q)\equiv\sum_{i=0}^{\infty}F_{0 \omega}^{(i)}(Q)=
\left(\frac{Q^2}{\mu^{2}}\right)^{\bar\gamma},
\end{eqnarray}
where $\bar\gamma$ is the BFKL anomalous dimension. 

Including coherence in the BFKL expressions, we get~\cite{Catani:1990yc,Marchesini:1995wr,Ciafaloni:1988ur,Catani:1991gu}:
\begin{eqnarray}
F_{\omega}(Q,\mu) &=& \Theta(Q-\mu) \nonumber\\
&+& \sum_{r=1}^{\infty}\int_{0}^{Q^{2}}
\prod_{i=1}^{r}\frac{d^{2}\mbox{\boldmath $q$}_{i}}{\pi q_{i}^{2}}dz_{i}
\frac{\bar{\alpha}_{S}}{z_{i}}z_{i}^{\omega}\Delta(z_{i},q_{i},k_{i})
\Theta(q_{i}-z_{i-1}q_{i-1}),
\end{eqnarray}
where $\Delta_{R}(z_{i},k_{i})$ is now the CCFM one
\begin{eqnarray}
\Delta(z_{i},q_{i},k_{i})=\exp\left[-\bar\alpha_{S}\ln\frac{1}{z_{i}}\ln
\frac{k_{i}^{2}}{z_{i}q_{i}^{2}}\right];~ ~ k_{i} > q_{i}.
\label{constraint}
\end{eqnarray}
For the first emission $q_{0}z_{0} = \mu$. The expansion of $F_{\omega}^{(r)}(Q)$ is 
\begin{eqnarray}
F_{\omega}^{(r)}(Q)=\sum_{n=r}^{\infty}\sum_{m=1}^{n}C^{(r)}(n,m;T)\frac{
\bar\alpha_{S}^{n}}{\omega^{2n-m}}.
\end{eqnarray}
The collinear cutoff is only needed in the first emission since 
subsequent emissions are regulated by angular ordering. 

The rates of emission of a number of gluons with transverse momentum larger 
than a scale $\mu_{R}$, with $\mu \ll \mu_{R}\ll Q$, plus any number of 
unresolved ones, were calculated in~\cite{Forshaw:1998uq} in the LLA to 
$\bar\alpha_S^3$. It was found that the jet rates both in the BFKL and CCFM 
approaches are the same:
\begin{eqnarray}
{\rm 0 ~jet} &=& \frac{(2\bar{\alpha}_{S})}{\omega}S
+\frac{(2\bar{\alpha}_{S})^{2}}{\omega^{2}}\left[\frac{S^{2}}{2}\right]
+\frac{(2\bar{\alpha}_{S})^{3}}{\omega^{3}}\left[\frac{S^{3}}{6}\right], \\  
{\rm 1~jet} &=& \frac{(2\bar{\alpha}_{S})}{\omega}T+\frac{(2\bar{\alpha}
_{S})^{2}}
{\omega^{2}}\left[TS-\frac{1}{2}T^{2}\right]\nonumber\\
&+&\frac{(2\bar{\alpha}_{S})^{3}}
{\omega^{3}}\left[\frac{1}{3}T^{3}-\frac{1}{2}T^{2}S+\frac{1}{2}TS^{2}\right],\\
{\rm 2~jet} &=& \frac{(2\bar{\alpha}_{S})^{2}}{\omega^{2}}
\left[T^{2}\right]+\frac{(2\bar{\alpha}_{S})^{3}}{\omega^{3}}
\left[T^{2}S-\frac{7}{6}T^{3}\right],\\
{\rm 3~jet} &=& \frac{(2\bar{\alpha}_{S})^{3}}{\omega^{3}}\left[T^{3}\right],
\end{eqnarray}
with $T=\ln(Q/\mu_R)$ and $S=\ln(\mu_R/\mu)$. This holds also to all orders 
in the coupling~\cite{Forshaw:1999yh} since a generating function for the jet 
multiplicity distribution was obtained in~\cite{Webber:1998we}:
\begin{eqnarray}
R^{(n ~{\rm jet})}_{\omega}(Q,\mu_R) = \frac{F^{(n {\rm ~jet})}_{\omega}(Q,\mu_R,\mu)}{F_{\omega}(Q,\mu)} = 
\frac{1}{n!} \left. \frac{\partial^n}{\partial u^n} R_{\omega}(u,T)\right|_{u=0},
\label{easy}
\end{eqnarray}
where the jet-rate generating function $R_{\omega}$ is given by
\begin{eqnarray}
R_{\omega}(u,T) = \exp{\left(-\frac{2 {\bar \alpha}_s}{\omega} T \right)} 
\left[1+(1-u)\frac{2 {\bar \alpha}_s}{\omega}T\right]^{\frac{u}{1-u}},
\end{eqnarray}
with the same generating function when coherence is included. The mean number
of jets and the mean square fluctuation are
\begin{eqnarray}
\langle n\rangle =
\left.\frac{\partial}{\partial u}R_{\omega}(u,T)\right|_{u=1} =
\frac{2 {\bar \alpha}_s}{\omega}T +\frac{1}{2} \left(\frac{2{\bar \alpha}_s}{\omega}T\right)^2 ,
\end{eqnarray}
\begin{eqnarray}
\langle n^2\rangle -\langle n\rangle^2 = \frac{2{\bar \alpha}_s}{\omega}T
+\frac{3}{2} \left(\frac{2{\bar \alpha}_s}{\omega}T\right)^2
+\frac{2}{3}\left(\frac{2{\bar \alpha}_s}{\omega}T\right)^3.
\end{eqnarray}
In~\cite{Ewerz:1999fn,Ewerz:1999tt} all subleading logarithms of 
$Q^{2}/\mu^{2}_{R}$ were included and the jet multiplicity in Higgs 
production at the LHC was 
found. It has also been shown that for any sufficiently inclusive 
observables the CCFM formalism leads to the same results as the BFKL equation~\cite{Salam:1999ft}. The implementation of CCFM in Monte Carlo event generators 
is discussed in, {\it e.g.},
~\cite{Dittmar:2005ed,Alekhin:2005dx,Alekhin:2005dy,Andersen:2006pg}. A 
numerical method suitable to investigate BFKL and CCFM in the NLLA in DIS is 
described in~Ref.~\cite{Andersen:2003an,Andersen:2003wy,Andersen:2004uj,Andersen:2004tt}.

\section{Beyond multi--Regge kinematics in the collinear region}
\label{collinear}

In~\cite{Salam:1998tj} multi--Regge kinematics was extended to include 
collinear contributions to all orders in the BFKL framework. 
In~\cite{Vera:2005jt} it was proved that this collinear region hides 
a very interesting structure in terms of double logarithms. A 
renormalization group (RG)--improved kernel was obtained which does not 
mix transverse with longitudinal momentum components. 

In ${\overline{\rm MS}}$ renormalisation the BFKL kernel in NLA 
reads~\cite{Fadin:1998py,Ciafaloni:1998gs}
\begin{eqnarray}
\label{ktKernel}
&&\hspace{-1.2cm}\int d^2 \vec{q}_2 \, {\cal K}\left(\vec{q}_1,\vec{q}_2\right) 
f\left({q}_2^2\right) = \nonumber\\
&&\hspace{-1cm}\int \frac{d^2 \vec{q}_2}{\left|{q}_1^2-{q}_2^2\right|} \left\{\left[{\bar \alpha}_s+{\bar \alpha}_s^2\left({\cal S}-\frac{\beta_0}{4 N_c}\ln{\left(
\frac{\left|{q}_1^2-{q}_2^2\right|^2}
{{\rm max}\left({q}_1^2,{q}_2^2\right)\mu^2}\right)}
\right)\right]\right.\nonumber\\
&&\hspace{2cm}\times\left(f\left({q}_2^2\right)-2 \frac{{\rm min}\left({q}_1^2,{q}_2^2\right)}{\left({q}_1^2+{q}_2^2\right)}f\left({q}_1^2\right)\right)\nonumber\\
&&\hspace{2cm}\left.\times-\frac{{\bar \alpha}_s^2}{4}\left({\cal T}\left({q}_1^2,{q}_2^2\right)+\ln^2{\left(\frac{{q}_1^2}{{q}_2^2}\right)}\right)f\left({q}_2^2\right)\right\}, 
\end{eqnarray}
with $\beta_0 = \left(11 N_c - 2 n_f\right)/3$,
${\cal S} = \left(4-\pi^2 +5 \beta_0/N_c \right)/12$. 
${\cal T} ({q}_1^2,{q}_2^2)$ can be found in~\cite{Fadin:1998py}. The 
action on the eigenfunctions at LLA is
\begin{eqnarray}
\label{nondiag}
\int d^2 \vec{q}_2 \, {\cal K} \left(\vec{q}_1,\vec{q}_2\right)
\left(\frac{{\bar \alpha}_s \left({q}_2^2\right)}{{\bar \alpha}_s \left({q}_1^2\right)}\right)^{-\frac{1}{2}}
\left(\frac{{q}_2^2}{{q}_1^2}\right)^{\gamma-1} = 
{\bar \alpha}_s \left({q}_1^2\right) \chi_0\left(\gamma\right)
+ {\bar \alpha}_s^2 \chi_1 \left(\gamma\right).
\end{eqnarray}
We have used 
\begin{eqnarray}
\label{LOeigen}
\chi_0 \left(\gamma\right) &=& 2 \psi(1) - \psi \left(\gamma\right)-
\psi \left(1-\gamma\right), 
\end{eqnarray}
\begin{eqnarray}
\label{chi1}
\chi_1 \left(\gamma\right) &=& {\cal S} \chi_0 \left(\gamma\right)
+ \frac{1}{4}\left(\psi''\left(\gamma\right)+ \psi''\left(1-\gamma\right)\right)
- \frac{1}{4}\left(\phi\left(\gamma\right)+ \phi\left(1-\gamma\right)\right)
+ \frac{3}{2} \zeta_3 \nonumber\\
&&\hspace{-1.9cm}-\frac{\pi^2 \cos{(\pi \gamma)}}{4 \sin^2(\pi \gamma)(1-2 \gamma)}
\left(3+\left(1+ \frac{n_f}{N_c^3}\right)\frac{(2+3\gamma(1-\gamma))}{(3-2\gamma)(1+2\gamma)}\right) - \frac{\beta_0}{8 N_c} \chi_0^2 \left(\gamma\right).
\end{eqnarray}
$\psi \left(\gamma \right) = \Gamma'\left(\gamma\right)/ 
\Gamma \left(\gamma\right)$ and
\begin{eqnarray}
\phi \left(\gamma\right) + \phi \left(1-\gamma\right) &=&\nonumber\\ 
&&\hspace{-3cm}\sum_{m=0}^{\infty} \left(\frac{1}{\gamma+m}+\frac{1}{1-\gamma+m}\right)
\left(\psi'\left(\frac{2+m}{2}\right)-\psi'\left(\frac{1+m}{2}\right)\right).
\end{eqnarray}
The poles in the collinear regions $\gamma = 0,1$ are 
\begin{eqnarray}
\chi_0 \left(\gamma\right) &\simeq& \frac{1}{\gamma} +
\left\{\gamma \rightarrow 1- \gamma\right\},\\
\chi_1 \left(\gamma \right) &\simeq& 
\frac{\rm a}{\gamma}+\frac{\rm b}{\gamma^2}-\frac{1}{2 \gamma^3} +
\left\{\gamma \rightarrow 1- \gamma\right\}.
\end{eqnarray}
where 
\begin{eqnarray}
\label{ab}
{\rm a} &=& \frac{5}{12}\frac{\beta_0}{N_c} -\frac{13}{36}\frac{n_f}{N_c^3}
-\frac{55}{36}, \, \, \,
{\rm b} ~=~ -\frac{1}{8}\frac{\beta_0}{N_c} -\frac{n_f}{6 N_c^3}
-\frac{11}{12}.
\end{eqnarray}
The cubic poles compensate for similar terms appearing when 
$s_0 = q_1 q_2$ is shifted to the DIS choice $s_0 = q_{1,2}^2$. 
Higher order terms beyond the NLLA, not compatible with RG evolution, 
are also generated by this change of scale. The truncation of the 
perturbative expansion is the reason why the gluon Green's function 
develops unphysical oscillations in the $q_1^2/q_2^2$ ratio.

To remove the most important terms in $\gamma$--space incompatible with RG 
evolution we simply perform the shift~\cite{Salam:1998tj}:
\begin{eqnarray}
\omega &=& {\bar \alpha}_s \left(1+\left({\rm a}+\frac{\pi^2}{6}\right){\bar \alpha}_s\right) \\
&&\hspace{1cm}\times\left(2 \psi(1)-\psi\left(\gamma+\frac{\omega}{2}-{\rm b}\,{\bar \alpha}_s \right)-\psi\left(1-\gamma+\frac{\omega}{2}-{\rm b}\,{\bar \alpha}_s \right)\right)\nonumber\\
&&\hspace{-0.6cm}+ {\bar \alpha}_s^2 \left(\chi_1 \left(\gamma\right) 
+\left(\frac{1}{2}\chi_0\left(\gamma\right)-{\rm b}\right)\left(\psi'(\gamma)+\psi'(1-\gamma)\right)-\left({\rm a}+\frac{\pi^2}{6}\right)\chi_0(\gamma)\right).
\nonumber
\end{eqnarray}
To solve this equation we consider the $\omega$--shift in the form
\begin{eqnarray}
&&{\omega \over {\bar \alpha}_s  \left(1+ {\rm A} {\bar \alpha}_s \right)}
= 2 \psi(1)-\psi\left(\gamma+\frac{\omega}{2}+ {\rm B} \,{\bar \alpha}_s\right)-\psi\left(1-\gamma+\frac{\omega}{2}+ {\rm B} \,{\bar \alpha}_s\right)
\nonumber\\
\label{generalshift}
&&\hspace{0.4cm}= \sum_{m=0}^{\infty} \left(\frac{1}{\gamma + m+\frac{\omega}{2}+ {\rm B} \,{\bar \alpha}_s}+
\frac{1}{1-\gamma+m+\frac{\omega}{2}+ {\rm B} \,{\bar \alpha}_s}-\frac{2}{m+1}\right).
\end{eqnarray}
We can now add all the approximated solutions at the different poles 
plus a substraction term to enforce convergence:
\begin{eqnarray}
\label{assumption}
\omega &=& \sum_{m=0}^{\infty} 
\left\{-(1+ 2 m + 2 \,{\rm B} \,{\bar \alpha}_s)+
\left|\gamma + m + {\rm B} \,{\bar \alpha}_s\right|
\left(1+\frac{2{\bar \alpha}_s \left(1+ {\rm A} {\bar \alpha}_s\right)}{\left(\gamma + m + {\rm B} \,{\bar \alpha}_s\right)^2}\right)^{\frac{1}{2}}\right. \nonumber\\
&&\hspace{-1cm}\left.+
\left|1-\gamma + m + {\rm B} \,{\bar \alpha}_s\right|
\left(1+\frac{2{\bar \alpha}_s \left(1+ {\rm A} {\bar \alpha}_s\right)}{\left(1-\gamma + m + {\rm B} \,{\bar \alpha}_s\right)^2}\right)^{\frac{1}{2}}
- \frac{2 {\bar \alpha}_s \left(1+ {\rm A} {\bar \alpha}_s\right)}{m+1}\right\}.
\end{eqnarray}
To match the original kernel at NLLA we set 
${\rm A} = {\rm a}$ and ${\rm B} = - {\rm b}$. The full NLLA scale invariant 
kernel without double counting terms then reads:
\begin{eqnarray}
\label{All-poles}
\omega &=& \bar{\alpha}_s \chi_0 (\gamma) + \bar{\alpha}_s^2 \chi_1 (\gamma) \\
&+& \left\{\sum_{m=0}^{\infty} \left[\left(\sum_{n=0}^{\infty}
\frac{(-1)^n (2n)!}{2^n n! (n+1)!}\frac{\left({\bar \alpha}_s+ {\rm a} \,{\bar \alpha}_s^2\right)^{n+1}}{\left(\gamma + m - {\rm b} \,{\bar \alpha}_s\right)^{2n+1}}\right) \right. \right. \nonumber\\&&\left.\left.-\frac{\bar{\alpha}_s}{\gamma + m} - \bar{\alpha}_s^2 \left(\frac{\rm a}{\gamma +m} + \frac{\rm b}{(\gamma + m)^2}-\frac{1}{2(\gamma+m)^3}\right)\right]+ \left\{\gamma \rightarrow 1-\gamma\right\}\right\}. \nonumber
\end{eqnarray}
\begin{figure}
\centering
\includegraphics[width=0.5\textwidth,angle=-90]{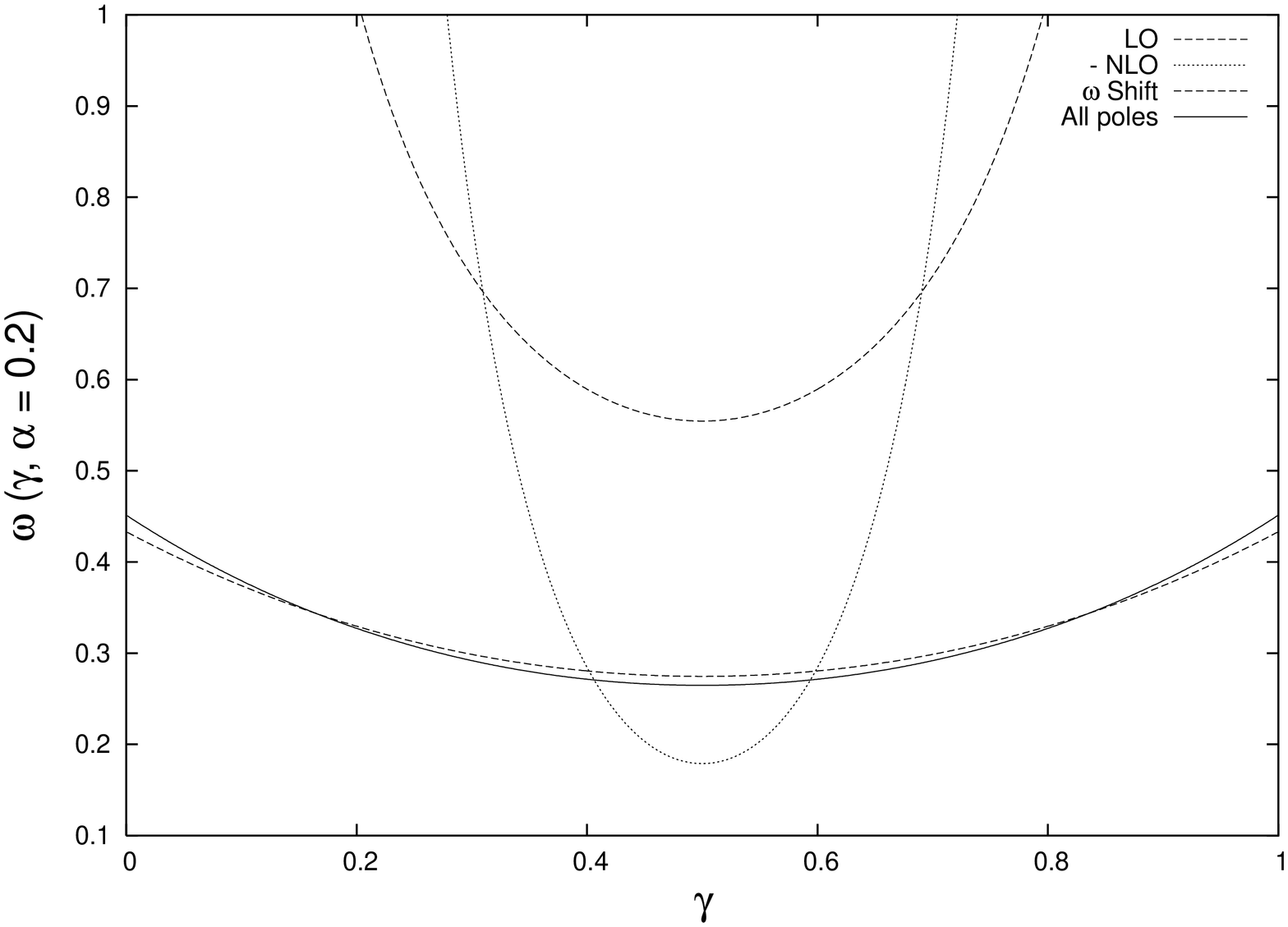}
\includegraphics[width=0.5\textwidth,angle=-90]{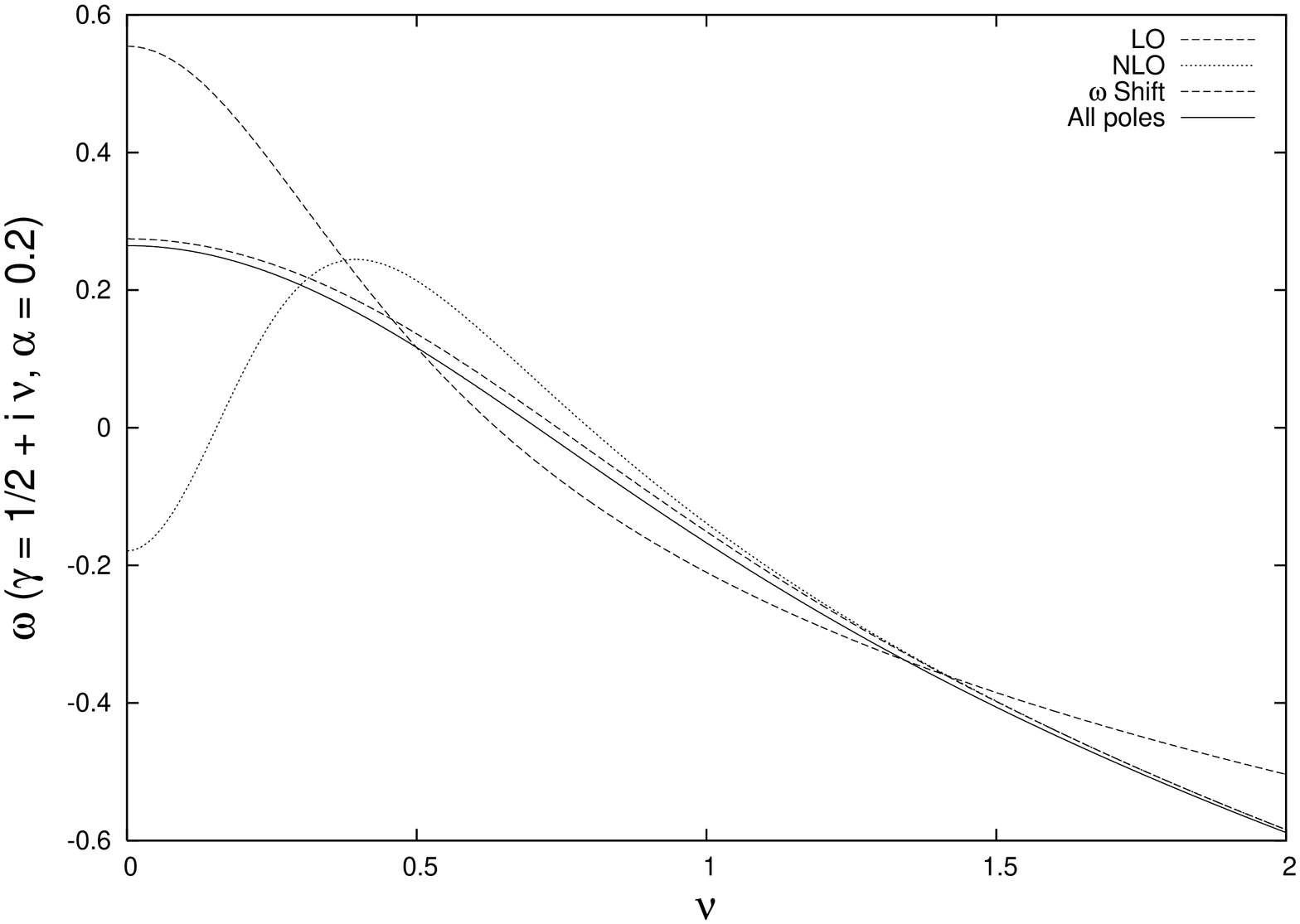}
\caption[]{The RG--improved kernel compared to its ``all--poles'' 
approximation together with the LO and NLO BFKL kernels.}
\label{FullFigure1}
\end{figure}
This result reproduces the $\omega$--shift very closely, 
see Fig.~\ref{FullFigure1}. It is very important to note that in 
Eq.~(\ref{All-poles}) the $\omega$--space is decoupled from the 
$\gamma$--representation. In~\cite{Vera:2005jt} an expression for the 
collinearly improved BFKL kernel which does not mix longitudinal with 
transverse degrees of freedom was obtained. To introduce the new kernel we 
only need to remove the term 
\begin{eqnarray}
\label{presc1}
-\frac{\bar{\alpha}_s^2}{4}\frac{1}{(\vec{q}-\vec{k})^2}
\ln^2\left({\frac{q^2}{k^2}}\right)
\end{eqnarray}
 in the emission part of the original kernel in the NLLA 
and replace it with
\begin{eqnarray}
\label{presc2}
\frac{1}{(\vec{q}-\vec{k})^2} \left\{\left(\frac{q^2}{k^2}\right)^{-{\rm b}{\bar \alpha}_s 
\frac{\left|k-q\right|}{k-q}}
\sqrt{\frac{2\left({\bar \alpha}_s+ {\rm a} \,{\bar \alpha}_s^2\right)}{\ln^2{\left(\frac{q^2}{k^2}\right)}}} 
J_1 \left(\sqrt{2\left({\bar \alpha}_s+ {\rm a} \,{\bar \alpha}_s^2\right) 
\ln^{2}{\left(\frac{q^2}{k^2}\right)}}\right) \right.\nonumber\\
&& \left.\hspace{-7cm}- {\bar \alpha}_s - {\rm a} \, {\bar \alpha}_s^2
+ {\rm b} \, {\bar \alpha}_s^2 \frac{\left|k-q\right|}{k-q}
\ln{\left(\frac{q^2}{k^2}\right)} \right\}.
\end{eqnarray}
For small differences between the $q^2$ and $k^2$ scales then 
\begin{eqnarray}
J_1 \left(\sqrt{2 {\bar \alpha}_s 
\ln^{2}{\left(\frac{q^2}{k^2}\right)}}\right) &\simeq& 
\sqrt{\frac{{\bar \alpha}_s}{2} \ln^{2}{\left(\frac{q^2}{k^2}\right)}},
\end{eqnarray}
and it does not change the ``Regge--like'' region. When 
the ratio of transverse momenta is large then 
\begin{eqnarray}
J_1 &\simeq& \left(\frac{2}{\pi^2 {\bar \alpha}_s  
\ln^{2}{\left(\frac{q^2}{k^2}\right)}}\right)^{\frac{1}{4}}
\cos{\left(\sqrt{2 {\bar \alpha}_s  
\ln^{2}{\left(\frac{q^2}{k^2}\right)}}-\frac{3\pi}{4}\right)},
\end{eqnarray}
removing the unphysical oscillations. This new kernel has been successfully 
applied 
to extend the region of applicability of NLLA BFKL calculations in the case 
of electroproduction of light vector mesons in~\cite{Caporale:2007vs}.

\section{Inclusive jet production at NLO}
\label{inclusivejet}

\begin{figure}[ht]
  \centering
  \includegraphics[width=8cm]{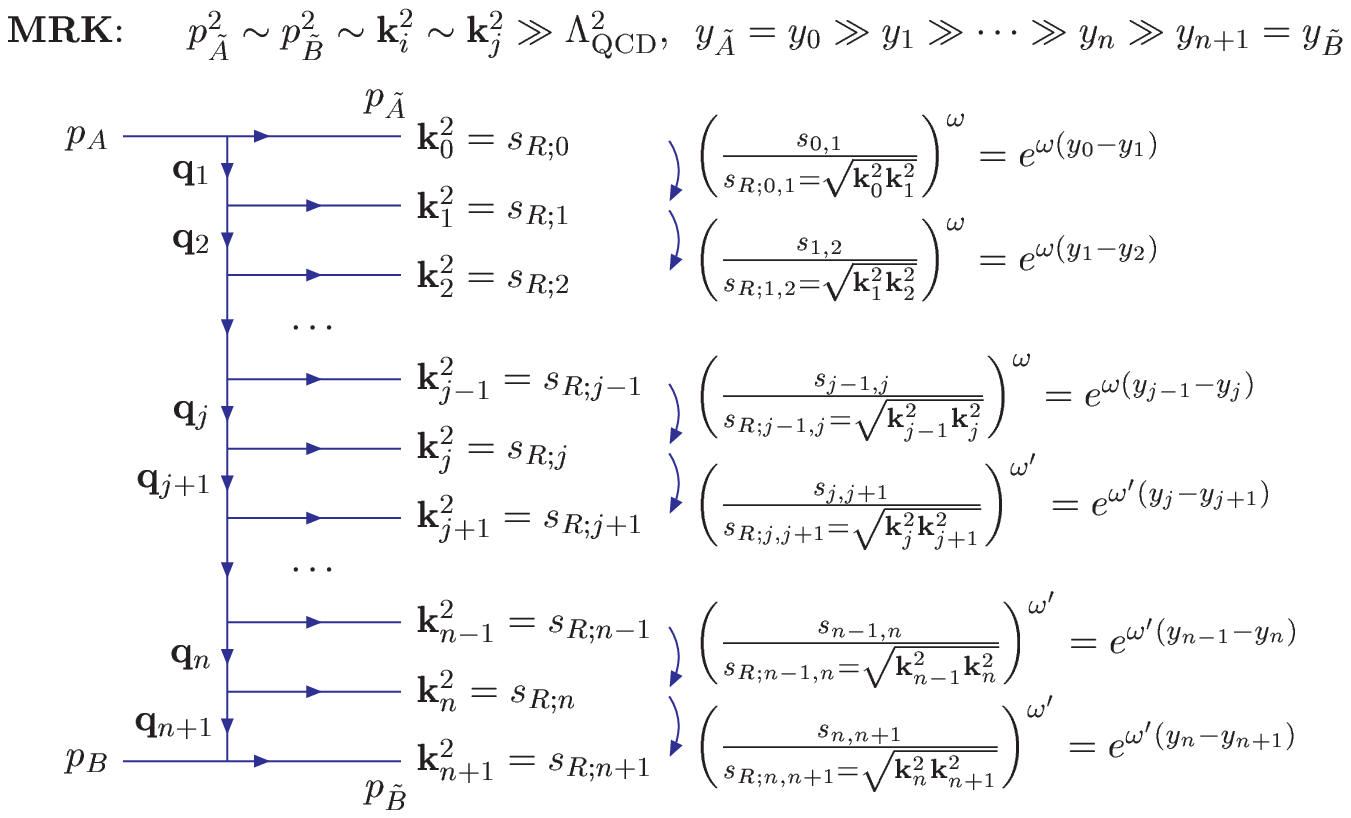}
  \caption{$2 \rightarrow 2 + (n-1) + {\rm jet}$ amplitude 
in the symmetric configuration with MRK. The produced jet has rapidity 
$y_J=y_j$ and transverse momentum ${\bf k}_J = {\bf k}_j$.}
  \label{SymmetricMRK}
\end{figure}

Now we discus the natural choice of $s_0$ when a hard jet is produced 
in the central region of rapidity~\cite{Bartels:2006hg}. Let us start with the symmetric case of 
$\gamma^*\gamma^*$ scattering with the virtualities of the two photons being 
large and of the same order. Here the rapidities of the emitted particles 
are the natural variables to characterize multijet production since all 
transverse momenta are of the same order. The rapidity difference between 
two emissions is 
\begin{eqnarray}
y_i - y_{i+1} &=& \ln{\frac{s_{i,i+1}}{\sqrt{{\bf k}_i^2 {\bf k}_{i+1}^2}}}
\end{eqnarray}
which supports the choice $s_{R;i,i+1}=\sqrt{{\bf k}_i^2{\bf k}_{i+1}^2}$ 
for the internal energy scales shown in Fig.~\ref{SymmetricMRK}.
In hadronic collisions MRK has to be modified to include 
evolution in the transverse momenta, since the momentum of the jet is 
larger than the typical transverse scale associated to the hadron. This can 
be done by changing the description of the evolution from one in terms of 
rapidities to another in terms of longitudinal momentum fractions of the
Reggeized gluons. 
Whereas in LO this change of scales has no consequences, in NLO accuracy it
leads to modifications, not only of the jet emission vertex but
also of the evolution kernels above and below the jet vertex. 
\begin{figure}
  \centering
  \includegraphics[width=7cm]{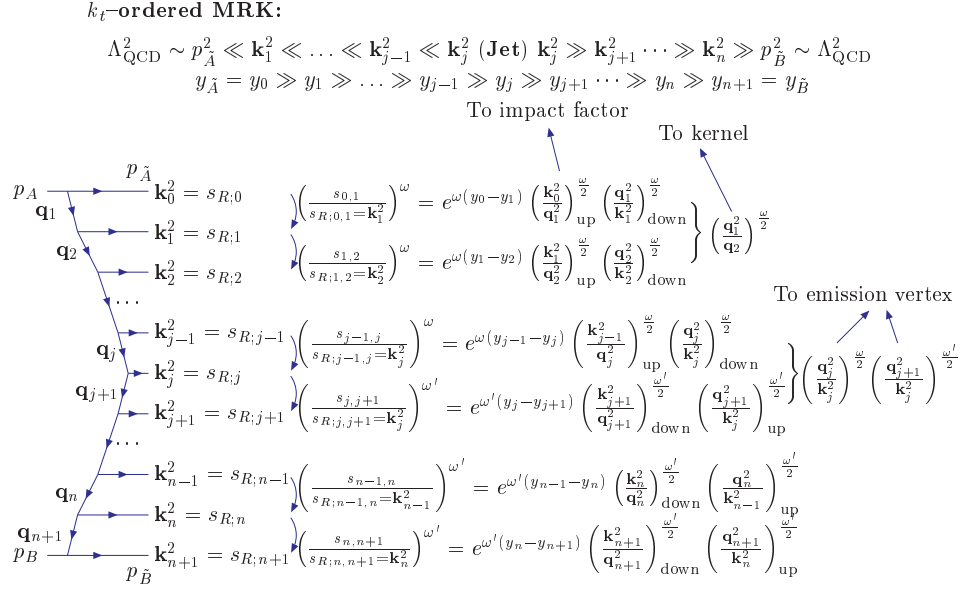}
  \caption{$2 \rightarrow 2 + (n-1) + {\rm jet}$ amplitude 
in the asymmetric configuration with $k_t$--ordered MRK.}
  \label{AsymmetricMRK}
\end{figure}

In more detail, we write the solution to the BFKL equation iteratively:
\begin{eqnarray}
\int d^2 \ka f_\omega(\ka,\qa) &=& 
\frac{1}{\omega} \sum_{j=1}^\infty 
\left[\prod_{i=1}^{j-1}\int d^2 \qi\frac{1}{\omega}\mathcal{K}(\qi,\qip)\right],
\label{eq:ggfiterando}
\end{eqnarray}
where $\qone = \ka$ and $\qj = \qa $, and think of one side of the 
evolution towards the hard scale using Fig.~\ref{AsymmetricMRK} as a guide. 
In the symmetric case the cross section contains the following evolution between particle $A$ and 
the jet:
\begin{eqnarray}
  \frac{d\sigma}{d^2 \kjet dy_J} &=& \int d^2 \qa \int d^2 \ka 
\frac{\Phi_A(\ka)}{2\pi\ka^2} \nonumber\\
&&\hspace{0.1cm}\times\int\frac{d\omega}{2\pi i} 
f_\omega(\ka,\qa)\left(\frac{s_{AJ}}{\sqrt{\ka^2\kjet^2}}\right)^\omega 
{\cal V}(\qa,\qb;\kjet,y_J)\dots
\end{eqnarray}

In the asymmetric situation where $\kjet^2\gg \ka^2$ the scale 
$\sqrt{\ka^2\kjet^2}$ should be replaced by $\kjet^2$. We then rewrite the 
term related to the choice of energy scale. Following 
Fig.~\ref{AsymmetricMRK} we take $\kj = \kjet$, 
${\bf k}_0 =  -\ka = -\qone$ and ${\bf q}_j = \qa$. It is convenient to 
introduce a chain of scale changes in every kernel:
\begin{eqnarray}
 \left(\frac{s_{AJ}}{\sqrt{\ka^2\kjet^2}}\right)^\omega &=& 
\left[\prod_{i=1}^{j}\left(\frac{\ki^2}{\kim^2}\right)^{\frac{\omega}{2}}\right]  \left(\frac{s_{AJ}}{\kjet^2}\right)^\omega, 
\end{eqnarray}
which can also be written in terms of the $t$--channel momenta as
\begin{eqnarray}
 \left(\frac{s_{AJ}}{\sqrt{\ka^2\kjet^2}}\right)^\omega &=& \left[\prod_{i=1}^{j-1}\left(\frac{\qip^2}{\qi^2}\right)^{\frac{\omega}{2}}\right]\left(\frac{\kjet^2}{\qa^2}\right)^{\frac{\omega}{2}}\left(\frac{s_{AJ}}{\kjet^2}\right)^\omega.
\end{eqnarray}
In this way we are changing the evolution from a difference in rapidity:
\begin{eqnarray}
\frac{s_{AJ}}{\sqrt{\ka^2\kjet^2}} &=& e^{y_{\tilde A}-y_J} 
\end{eqnarray}
to the inverse of the longitudinal momentum fraction, {\it i.e.}
\begin{eqnarray}
\frac{s_{AJ}}{\kjet^2} &=& \frac{1}{\alpha_J}.
\label{jetinverselong}
\end{eqnarray}
This shift in scales affects the expression for the cross section:
\begin{eqnarray}
\frac{d\sigma}{d^2 \kjet dy_J} &=& 
\int\frac{d\omega}{2\pi i \, \omega}\sum_{j=1}^\infty 
\left[\prod_{i=1}^{j}\int d^2 \qi\right] \frac{\Phi_A(\qone)}{2\pi\qone^2}
\nonumber\\
&&\hspace{-2.5cm}\times\left[\prod_{i=1}^{j-1}\left(\frac{\qip^2}{\qi^2}\right)^{\frac{\omega}{2}}\frac{1}{\omega}\mathcal{K}(\qi,\qip)\right]\left(\frac{\kjet^2}{\qa^2}\right)^{\frac{\omega}{2}}
{\cal V}(\qa,\qb;\kjet,y_J)\left(\frac{s_{AJ}}{\kjet^2}\right)^\omega\ldots
\end{eqnarray}
These changes can be absorbed at NLO in the kernels and 
impact factors. The impact factors get one contribution, as can be seen in 
Fig.~\ref{AsymmetricMRK}:
\begin{eqnarray}
\label{newimpactfactor}
 \widetilde{\Phi}(\ka)&=& \Phi(\ka) -\frac{1}{2}{\ka^2}\int d^2 {\bf q} 
\frac{\Phi^{(B)}({\bf q})}{{\bf q}^2}\mathcal{K}^{(B)}({\bf q},\ka)
\ln\frac{{\bf q}^2}{\ka^2}.
\end{eqnarray}
The kernels in the evolution receive a double contribution from the different 
energy scale choices of both the incoming and outgoing Reggeons (see 
Fig.~\ref{AsymmetricMRK}). This amounts to the following correction:
\begin{eqnarray}
\label{newkernel}
  \widetilde{\mathcal{K}}(\qone,\qtwo) &=& \mathcal{K}(\qone,\qtwo)
-\frac{1}{2}\int d^2 {\bf q} \, \mathcal{K}^{(B)}(\qone,{\bf q}) 
\, \mathcal{K}^{(B)}({\bf q},\qtwo)\ln\frac{{\bf q}^2}{\qtwo^2}.
\end{eqnarray}
There is a different type of term in the case of the emission vertex 
where the jet is defined. This correction has also two contributions 
originated at the two different evolution chains from the hadrons $A$ and 
$B$:
\begin{eqnarray}
\label{newemissionvertex}
  \widetilde{\cal V}(\qa,\qb) &=& {\cal V}(\qa,\qb)
-\frac{1}{2}\int d^2 {\bf q} \,  
\mathcal{K}^{(B)}(\qa,{\bf q}) {\cal V}^{(B)}({\bf q},\qb)
\ln\frac{{\bf q}^2}{({\bf q}-\qb)^2}\nonumber\\
&&\hspace{1cm}-\frac{1}{2}\int d^2 {\bf q} \, {\cal V}^{(B)}(\qa,{\bf q}) \,
\mathcal{K}^{(B)}({\bf q},\qb)\ln\frac{{\bf q}^2}{(\qa-{\bf q})^2}.
\end{eqnarray}
The final expression for the cross section in the asymmetric case is
\begin{eqnarray}
  \frac{d\sigma}{d^2 \kjet dy_J} &=& \int d^2 \qa\int d^2 \ka 
\frac{\widetilde{\Phi}_A(\ka)}{2\pi\ka^2}\nonumber\\
&&\hspace{1cm}\times \int\frac{d\omega}{2\pi i}
\tilde{f}_\omega(\ka,\qa)\left(\frac{s_{AJ}}{\kjet^2}\right)^\omega 
\widetilde{\cal V}(\qa,\qb;\kjet,y_J)\ldots
\end{eqnarray}
It is interesting to discuss the NLO unintegrated gluon density in this 
context. It is defined by
\begin{equation}
  g(x,{\bf k}) = \int d^2 {\bf q}\frac{\widetilde{\Phi}_P({\bf q})}{2\pi{\bf q}^2}\int\frac{d\omega}{2\pi i}\tilde{f}_\omega({\bf k},{\bf q})\, x^{-\omega},
\end{equation}
where the gluon Green's function ${\tilde f}_\omega$ is the solution to a new 
BFKL equation with the modified kernel which includes the energy shift at NLO:
\begin{equation}
  \omega \tilde{f}_\omega(\ka,\qa) = \del{\ka-\qa} +
\int d^2{\bf q} \, \widetilde{\mathcal{K}}(\ka,{\bf q}) \,
\tilde{f}_\omega({\bf q},\qa).
\end{equation}
The unintegrated gluon distribution then follows the evolution equation 
\begin{equation}
  \frac{\partial g(x,\qa)}{\partial\ln 1/x} 
= \int d^2 {\bf q} \, \widetilde{\mathcal{K}} (\qa,{\bf q}) \, g(x,{\bf q}).
\end{equation}
Finally, taking into account the evolution from the other hadron, the 
differential cross section reads
 \begin{equation}
  \frac{d\sigma}{d^2 \kjet dy_J} = \int d^2 \qa \int d^2 \qb 
\, g(x_a,\qa) \, g(x_b,\qb) \, \widetilde{\cal V}(\qa,\qb;\kjet,y_J).
\label{ppfinal} 
\end{equation}
It is worth mentioning that the proton impact factor contains 
non--perturbative physics which can only be modeled by, {\it e.g.}
\begin{eqnarray}
\Phi_P({\bf q}) &\sim& (1-x)^{p_1} x^{- p_2} 
\left(\frac{{\bf q}^2}{{\bf q}^2+Q_0^2}\right)^{p_3},
\label{eq:protonIF}
\end{eqnarray}
where $p_i$ are positive free parameters and $Q_0^2$ representing a momentum
scale of the order of the confinement scale.  

\section{Azimuthal angle decorrelations in Mueller--Navelet jets at hadron 
colliders}
\label{conformal}

In~\cite{Vera:2006un} azimuthal angle decorrelations in inclusive dijet cross 
sections were studied analytically including the BFKL kernel in the NLLA 
while keeping the jet vertices at leading order. The angular decorrelation for 
jets with a wide separation in rapidity decreases when NLO effects are 
included.

BFKL effects should dominate in observables with a large 
center--of--mass energy, and two large and similar transverse scales. This is 
the case of the inclusive hadroproduction of two jets with large and similar 
transverse momenta  and a large relative separation in rapidity, Y. These are 
the so--called Mueller--Navelet jets, first proposed in 
Ref.~\cite{Mueller:1986ey}. A rise with Y in the partonic cross section 
was predicted in agreement with the LLA hard Pomeron intercept. At 
hadronic level Mueller--Navelet jets are produced in a region of fast 
falling of the parton distributions, reducing this rise. BFKL enhances soft 
real emission as Y increases reducing the angular correlation. This was 
investigated in the LLA in 
Ref.~\cite{DelDuca:1993mn,DelDuca:1994ng,Stirling:1994zs}. The decorrelation 
lies quite below the experimental 
data~\cite{Abachi:1996et,Abbott:1999ai,Abbott:1997nf,Abe:1996mj} at the 
Tevatron. 

We now investigate the cross section parton + parton $\rightarrow$ jet + 
jet + soft emission, with the two jets having transverse momenta $\vec{q}_1$ 
and $\vec{q}_2$ and with a relative rapidity separation Y. The differential 
partonic cross section is
\begin{eqnarray}
\frac{d {\hat \sigma}}{d^2\vec{q}_1 d^2\vec{q}_2} &=& \frac{\pi^2 {\bar \alpha}_s^2}{2} 
\frac{f \left(\vec{q}_1,\vec{q}_2,{\rm Y}\right)}{q_1^2 q_2^2},
\end{eqnarray}
We work with the Mellin transform:
\begin{eqnarray}
f \left(\vec{q}_1,\vec{q}_2,{\rm Y}\right) &=& \int \frac{d\omega}{2 \pi i} e^{\omega {\rm Y}} f_\omega \left(\vec{q}_1,\vec{q}_2\right).
\end{eqnarray}
The solution to the BFKL equation in the LLA is
\begin{eqnarray}
f_\omega \left(\vec{q}_1,\vec{q}_2\right) = \frac{1}{2 \pi^2} \sum_{n = -\infty}^\infty
\int_{-\infty}^\infty d \nu \,{\left(q_1^2\right)}^{-i \nu -\frac{1}{2}} {\left(q_2^2\right)}^{i \nu -\frac{1}{2}} \frac{e^{i n \left(\theta_1-\theta_2\right)}}{\omega - {\bar \alpha}_s \chi_0 \left(\left|n\right|,\nu\right)} 
\end{eqnarray}
with
\begin{eqnarray}
\chi_0 \left(n,\nu\right) &=& 2 \psi \left(1\right) - \psi \left(\frac{1}{2}+ i \nu + \frac{n}{2}\right) - \psi\left(\frac{1}{2}- i \nu +\frac{n}{2}\right), 
\end{eqnarray}
The BFKL equation for non--zero momentum transfer is of Schr{\"o}dinger--like 
type with a holomorphically separable Hamiltonian. Both the holomorphic and 
antiholomorphic sectors are invariant under spin zero M{\"o}bius 
transformations with eigenfunctions carrying a conformal weight of the form 
$\gamma = \frac{1}{2} + i \nu + \frac{n}{2} $. In the principal series of the 
unitary representation $\nu$ is real and $\left|n\right|$ the integer 
conformal spin~\cite{Lipatov:1985uk}. Hence, extracting information 
about $n$ is equivalent to proving the conformal structure of high energy QCD.
 
We now integrate over the phase space of the two emitted gluons together with 
some general jet vertices, {\it i.e.}
\begin{eqnarray}
{\hat \sigma} \left(\alpha_s, {\rm Y},p^2_{1,2}\right) =
\int d^2{\vec{q}_1} \int d^2{\vec{q}_2} \,
\Phi_{\rm jet_1}\left(\vec{q}_1,p_1^2\right)
\,\Phi_{\rm jet_2}\left(\vec{q}_2,p_2^2\right)\frac{d {\hat \sigma}}{d^2\vec{q}_1 d^2\vec{q}_2}.
\end{eqnarray}
In the jet vertices only leading--order terms are kept:
\begin{eqnarray}
\Phi_{\rm jet_i}^{(0)} \left(\vec{q},p_i^2\right)&=& \theta \left(q^2-p_i^2\right),
\end{eqnarray}
where $p_i^2$ corresponds to a resolution scale for the gluon jet. To extend 
this analysis it is needed to use the NLO jet 
vertices in Ref.~\cite{Bartels:2001ge,Bartels:2002yj} where 
the definition of a jet is much more complex than here. We can now write
\begin{eqnarray}
{\hat \sigma} &=& \frac{\pi^2 {\bar \alpha}_s^2}{2} \int d^2{\vec{q}_1} \int d^2{\vec{q}_2} \,
\frac{\Phi_{\rm jet_1}^{(0)}\left(\vec{q}_1,p_1^2\right)}{q_1^2}
\,\frac{\Phi_{\rm jet_2}^{(0)}\left(\vec{q}_2,p_2^2\right)}{q_2^2}
f \left(\vec{q}_1,\vec{q}_2,{\rm Y}\right). 
\end{eqnarray}
In a transverse momenta operator representation:
\begin{eqnarray}
\left< \vec{q}\right|\left.\nu,n\right> &=& \frac{1}{\pi \sqrt{2}} 
\left(q^2\right)^{i \nu -\frac{1}{2}} \, e^{i n \theta}, 
\label{eignfns}
\end{eqnarray}
the action of the NLO kernel, calculated in Ref.~\cite{Kotikov:2000pm}, is 
\begin{eqnarray}
{\hat K} \left|\nu,n\right> &=& \left\{\frac{}{}{\bar \alpha}_s \, \chi_0\left(\left|n\right|,\nu\right) + {\bar \alpha}_s^2 \, \chi_1\left(\left|n\right|,\nu\right) \right.\nonumber\\
&&\left.\hspace{-2cm}+\,{\bar \alpha}_s^2 \,\frac{\beta_0}{8 N_c}\left[2\,\chi_0\left(\left|n\right|,\nu\right) \left(i \frac{\partial}{\partial \nu}+ \log{\mu^2}\right)+\left(i\frac{\partial}{\partial \nu}\chi_0\left(\left|n\right|,\nu\right)\right)\right]\right\} \left|\nu,n\right>,
\label{opKernel}
\end{eqnarray}
where $\chi_1$, for a general conformal spin, reads 
\begin{eqnarray}
\chi_1\left(n,\gamma \right) &=& {\cal S} \chi_0 \left(n, \gamma\right)
+ \frac{3}{2} \zeta\left(3\right) - \frac{\beta_0}{8 N_c}\chi_0^2\left(n,\gamma\right)\nonumber\\
&+&\frac{1}{4}\left[\psi''\left(\gamma+\frac{n}{2}\right)+\psi''\left(1-\gamma+\frac{n}{2}\right)-2 \,\phi\left(n,\gamma\right)-2 \,\phi\left(n,1-\gamma\right)\right]\nonumber\\
&-&\frac{\pi^2 \cos{\left(\pi \gamma\right)}}{4 \sin^2\left(\pi \gamma\right)\left(1-2\gamma\right)}\left\{\left[3+\left(1+\frac{n_f}{N_c^3}\right)\frac{2+3\gamma\left(1-\gamma\right)}{\left(3-2\gamma\right)\left(1+2\gamma\right)}\right]\delta_{n 0}\right.\nonumber\\
&&\left.\hspace{2cm}-\left(1+\frac{n_f}{N_c^3}\right)\frac{\gamma\left(1-\gamma\right)}{2\left(3-2\gamma\right)\left(1+2\gamma\right)}\delta_{n 2}\right\},
\end{eqnarray}
with ${\cal S}=\left(4-\pi^2+5 \beta_0/N_c\right)/12$, 
$\beta_0 = (11 N_c-2 n_f)/3$. $\phi$ can be found in~\cite{Kotikov:2000pm}.

\begin{figure}
\centering
\includegraphics[width=0.5\textwidth,angle=-90]{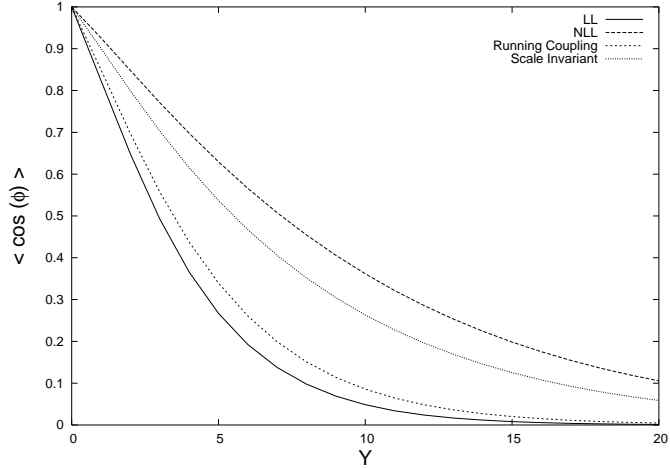}
\caption[]{Dijet azimuthal angle correlation as a function of the rapidity separation.}
\label{Cos1Y}
\end{figure}

The jet vertices on the basis in Eq.~(\ref{eignfns}) are:
\begin{eqnarray}
\int d^2{\vec{q}}\,\frac{\Phi_{\rm jet_1}^{(0)}\left(\vec{q},p_1^2\right)}{q^2} \left<\vec{q}\right.\left|\nu,n\right> = 
\frac{1}{\sqrt{2}}\frac{1}{\left(\frac{1}{2}-i \nu\right)}\left(p_1^2\right)^{i \nu - \frac{1}{2}} \delta_{n,0} \equiv c_1\left(\nu\right) \delta_{n,0},
\label{IFproj}
\end{eqnarray}
with the $c_2\left(\nu\right)$ projection of $\Phi_{\rm jet_2}^{(0)}$ on 
$\left<n,\nu\right|\left.\vec{q}\right>$ being the complex conjugate 
of~(\ref{IFproj}) with $p_1^2$ being replaced by $p_2^2$. The cross section 
now reads
\begin{eqnarray}
{\hat \sigma} &=&
\frac{\pi^2 {\bar \alpha}_s^2}{2} \sum_{n=-\infty}^\infty \int_{-\infty}^\infty d \nu 
\,e^{{\bar \alpha}_s \chi_0\left(\left|n\right|,\nu\right) {\rm Y}} 
c_1\left(\nu\right) c_2\left(\nu\right) \delta_{n,0} \Bigg\{1+{\bar \alpha}_s^2 \, {\rm Y} \\ 
&&\hspace{-0.6cm}\times\left[\chi_1\left(\left|n\right|,\nu\right)+\frac{\beta_0}{4 N_c} \left(\log{(\mu^2)}+ \frac{i}{2} \frac{\partial}{\partial \nu}\log{\left(\frac{c_1\left(\nu\right)}{c_2\left(\nu\right)}\right)}+ \frac{i}{2} \frac{\partial}{\partial \nu}\right)\chi_0\left(\left|n\right|,\nu\right)\right]\Bigg\}. \nonumber
\label{logdercross}
\end{eqnarray}
For the LO jet vertices the logarithmic derivative in Eq.~(\ref{logdercross}) 
is 
\begin{eqnarray}
- i \frac{\partial}{\partial \nu}\log{\left(\frac{c_1\left(\nu\right)}{c_2\left(\nu\right)}\right)} &=& \log{\left(p_1^2p_2^2\right)}+ \frac{1}{\frac{1}{4}+\nu^2}.
\end{eqnarray}
If $\phi = \theta_1-\theta_2 - \pi$, in the case of two equal resolution 
momenta, $p_1^2 = p_2^2 \equiv p^2$, the angular differential cross section 
can be expressed as
\begin{eqnarray}
\frac{d {\hat \sigma}\left(\alpha_s, {\rm Y},p^2\right)}{d \phi}  &=&
\frac{\pi^3 {\bar \alpha}_s^2}{2 p^2} \frac{1}{2 \pi}\sum_{n=-\infty}^\infty 
e^{i n \phi} {\cal C}_n \left({\rm Y}\right),
\end{eqnarray}
with
\begin{eqnarray}
{\cal C}_n \left({\rm Y}\right) =
\int_{-\infty}^\infty \frac{d \nu}{2 \pi}\frac{e^{{\bar \alpha}_s \left(p^2\right){\rm Y} \left(\chi_0\left(\left|n\right|,\nu\right)+{\bar \alpha}_s  \left(p^2\right) \left(\chi_1\left(\left|n\right|,\nu\right)-\frac{\beta_0}{8 N_c} \frac{\chi_0\left(\left|n\right|,\nu\right)}{\left(\frac{1}{4}+\nu^2\right)}\right)\right)}}{\left(\frac{1}{4}+\nu^2\right)}.
\label{Cn}
\end{eqnarray}
$n=0 $ governs the energy dependence of the cross section:
\begin{eqnarray}
{\hat \sigma}\left(\alpha_s, {\rm Y},p^2\right) &=& 
\frac{\pi^3 {\bar \alpha}_s^2}{2 p^2} \, {\cal C}_0 \left({\rm Y}\right).
\end{eqnarray}
In the plots we take $p = 30 \, {\rm GeV}$, $n_f = 4$ and 
$\Lambda_{\rm QCD} = 0.1416$ GeV. The $n = 0$ coefficient is directly related 
to the normalized cross section
\begin{eqnarray}
\frac{{\hat \sigma} \left({\rm Y}\right)}{{\hat \sigma} \left(0\right)} 
&=& \frac{{\cal C}_0 \left({\rm Y}\right)}{{\cal C}_0\left(0\right)}. 
\end{eqnarray}
\begin{figure}
\centering
\includegraphics[width=0.5\textwidth,angle=-90]{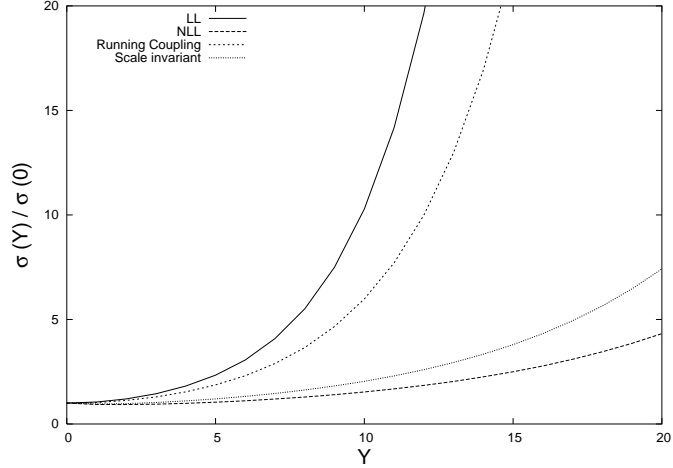}
\caption[]{Partonic cross section growth with the rapidity separation of the dijets.}
\label{SectionconY}
\end{figure}
The rise with Y of this observable is shown in 
Fig.~\ref{SectionconY}. Clearly the NLL intercept is very much reduced with 
respect to the LL case. The remaining coefficients with $n \geq 1$ all 
decrease with Y. Because of this the angular correlations also 
diminish as the rapidity interval between the jets gets larger. This point 
can be studied in detail using the mean values 
\begin{eqnarray}
\left<\cos{\left( m \phi \right)}\right> &=& \frac{{\cal C}_m \left({\rm Y}\right)}{{\cal C}_0\left({\rm Y}\right)}.
\end{eqnarray}
$\left<\cos{\left(\phi\right)}\right>$ is calculated in Fig.~\ref{Cos1Y}. The 
NLL effects decrease the azimuthal angle decorrelation. This is the case for 
the running of the coupling and also for the scale invariant terms. This is 
encouraging from the phenomenological point of view given that the data at the 
Tevatron typically have lower decorrelation than predicted by LLA BFKL or LLA 
with running coupling. 
The difference in the decorrelation between LLA and NLLA is driven by the $n=0$ 
conformal spin since the ratio
\begin{eqnarray}
\frac{\left<\cos{\left(\phi \right)}\right>^{\rm NLLA}}{\left<\cos{\left(\phi \right)}\right>^{\rm LLA}} &=& \frac{{\cal C}_1^{\rm NLLA} \left({\rm Y}\right)}{{\cal C}_0^{\rm NLLA}\left({\rm Y}\right)}\frac{{\cal C}_0^{\rm LLA} \left({\rm Y}\right)}{{\cal C}_1^{\rm LLA}\left({\rm Y}\right)},
\end{eqnarray}
is always close to one
\begin{eqnarray}
1.2 > \frac{{\cal C}_1^{\rm NLLA} \left({\rm Y}\right)}{{\cal C}_1^{\rm LLA}\left({\rm Y}\right)} > 1.
\end{eqnarray}
\begin{figure}
\centering
\includegraphics[width=0.5\textwidth,angle=-90]{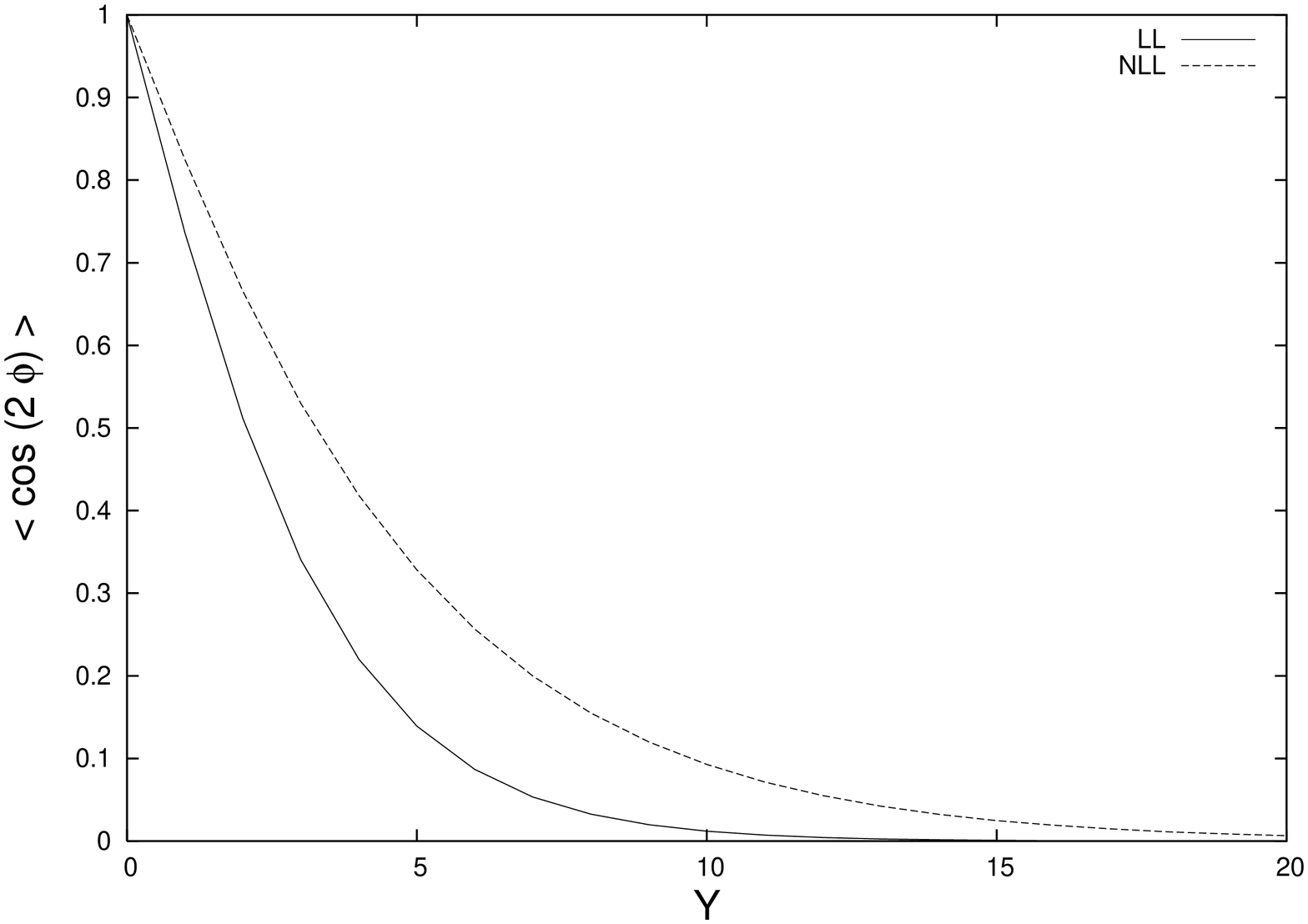}
\includegraphics[width=0.5\textwidth,angle=-90]{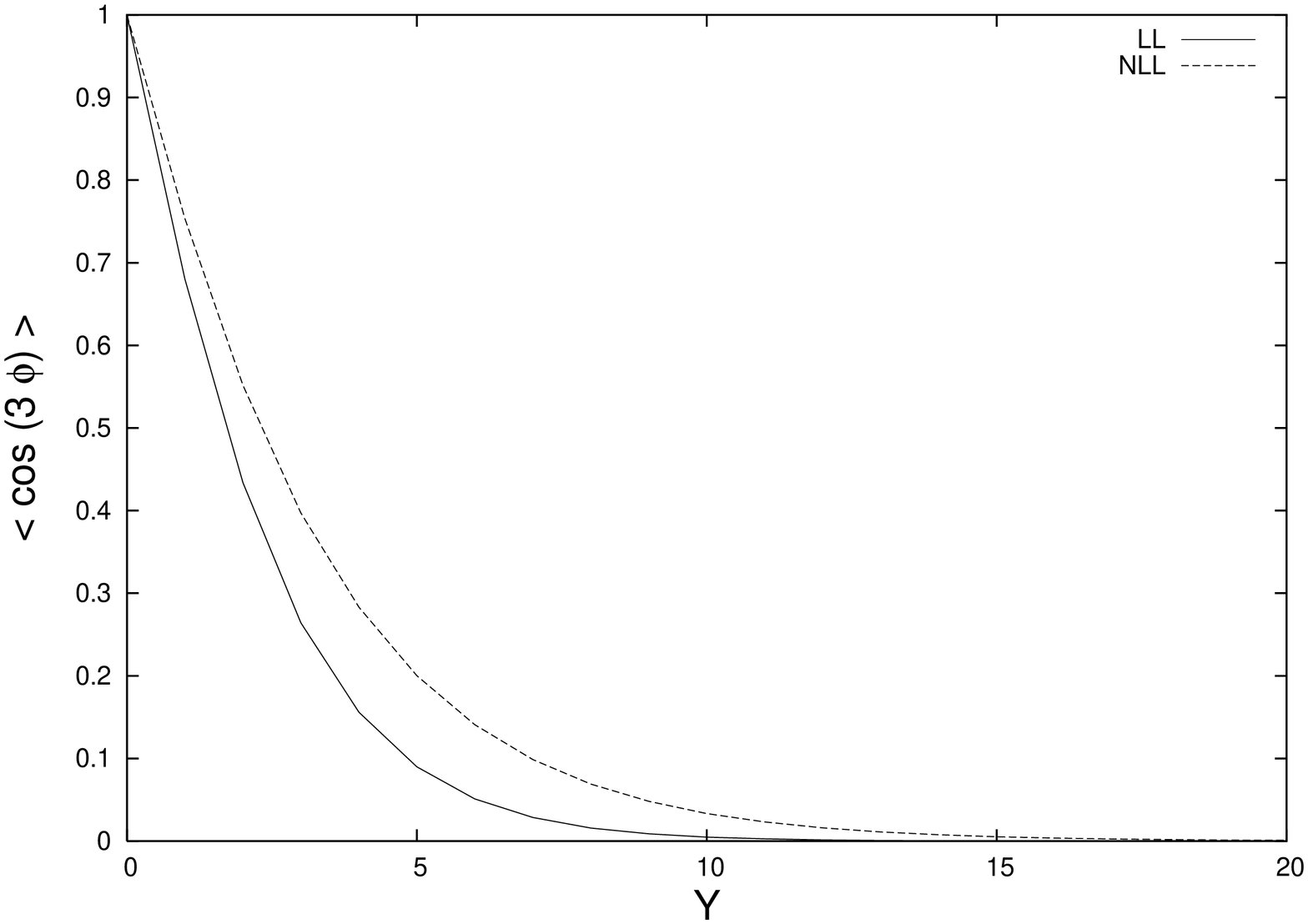}
\caption[]{Dijet azimuthal angle decorrelation as a function of their separation in rapidity.}
\label{Cos23Y}
\end{figure}
This is a consequence of the good convergence in terms of asymptotic 
intercepts of the NLLA BFKL calculation for conformal spins larger than zero. 
For completeness the $m=2,3$ cases for 
$\left<\cos{\left(m \phi\right)}\right>$ are shown in Fig.~\ref{Cos23Y}. 
These distributions test the structure of the 
higher conformal spins. The methods of this section have been applied to 
phenomenology of dijets at the Tevatron and the LHC 
in~\cite{Vera:2007kn,Marquet:2007xx}, and to the production of forward jets 
in DIS at HERA in~\cite{Vera:2007dr}.


\begin{thebibliography}{99}

\bibitem{BFKL1}  
L.~N.~Lipatov, 
Sov.\ J.\ Nucl.\ Phys.\  {\bf 23} (1976) 338.

\bibitem{BFKL2}
V.~S.~Fadin, E.~A.~Kuraev, L.~N.~Lipatov, 
Phys.\ Lett.\  B {\bf 60} (1975) 50.

\bibitem{BFKL3}
E.~A.~Kuraev, L.~N.~Lipatov, V.~S.~Fadin,
Sov.\ Phys.\ JETP {\bf 44} (1976) 443.

\bibitem{BFKL4}
E.~A.~Kuraev, L.~N.~Lipatov, V.~S.~Fadin,
Sov.\ Phys.\ JETP {\bf 45} (1977) 199.

\bibitem{BFKL5}
I.~I.~Balitsky, L.~N.~Lipatov, 
Sov.\ J.\ Nucl.\ Phys.\  {\bf 28} (1978) 822. 

\bibitem{Catani:1990yc}
S.~Catani, F.~Fiorani and G.~Marchesini, 
Phys.\ Lett.\  {\bf B234}, 339 (1990).

\bibitem{Marchesini:1995wr}
G.~Marchesini, 
Nucl.\ Phys.\  {\bf B445}, 49 (1995).

\bibitem{Ciafaloni:1988ur}
M.~Ciafaloni, 
Nucl.\ Phys.\  {\bf B296}, 49 (1988).

\bibitem{Catani:1991gu}
S.~Catani, F.~Fiorani, G.~Marchesini and G.~Oriani, 
Nucl.\ Phys.\  {\bf B361}, 645 (1991).

\bibitem{Forshaw:1998uq}
J.~R.~Forshaw and A.~Sabio Vera, 
Phys.\ Lett.\  {\bf B440} (1998) 141.

\bibitem{Forshaw:1999yh}
J.~R.~Forshaw, A.~Sabio Vera and B.~R.~Webber, 
J.\ Phys.\ G {\bf G25} (1999) 1511.

\bibitem{Webber:1998we}
B.~R.~Webber, Phys.\ Lett.\  {\bf B444} (1998) 81.

\bibitem{Ewerz:1999fn}
C.~Ewerz and B.~R.~Webber, JHEP {\bf 9904} (1999) 022.

\bibitem{Ewerz:1999tt}
C.~Ewerz and B.~R.~Webber, JHEP {\bf 9908} (1999) 019.

\bibitem{Salam:1999ft}
G.~P.~Salam, JHEP {\bf 9903} (1999) 009.

\bibitem{Dittmar:2005ed}
  M.~Dittmar {\it et al.},  arXiv:hep-ph/0511119.

\bibitem{Alekhin:2005dx}
  S.~Alekhin {\it et al.}, arXiv:hep-ph/0601012.

\bibitem{Alekhin:2005dy}
  S.~Alekhin {\it et al.}, arXiv:hep-ph/0601013.

\bibitem{Andersen:2006pg}
  J.~R.~Andersen {\it et al.}  [Small x Collaboration],
  Eur.\ Phys.\ J.\  C {\bf 48} (2006) 53.

\bibitem{Andersen:2003an}
  J.~R.~Andersen and A.~Sabio Vera,
  Phys.\ Lett.\  B {\bf 567} (2003) 116.

\bibitem{Andersen:2003wy}
  J.~R.~Andersen and A.~Sabio Vera,
  Nucl.\ Phys.\  B {\bf 679} (2004) 345.

\bibitem{Andersen:2004uj}
  J.~R.~Andersen and A.~Sabio Vera,
  Nucl.\ Phys.\  B {\bf 699} (2004) 90.

\bibitem{Andersen:2004tt}
  J.~R.~Andersen and A.~Sabio Vera,
  JHEP {\bf 0501} (2005) 045.


\bibitem{Salam:1998tj}
  G.~P.~Salam, JHEP {\bf 9807}, 019 (1998).

\bibitem{Vera:2005jt}
  A.~Sabio Vera, Nucl.\ Phys.\  B {\bf 722} (2005) 65.

\bibitem{Fadin:1998py}
  V.~S.~Fadin and L.~N.~Lipatov, Phys.\ Lett.\  B {\bf 429} (1998) 127.

\bibitem{Ciafaloni:1998gs}
  M.~Ciafaloni and G.~Camici, Phys.\ Lett.\  B {\bf 430} (1998) 349.

\bibitem{Caporale:2007vs}
  F.~Caporale, A.~Papa and A.~Sabio Vera,
  Eur.\ Phys.\ J.\  C {\bf 53} (2008) 525.

\bibitem{Bartels:2006hg}
  J.~Bartels, A.~Sabio Vera and F.~Schwennsen,
  JHEP {\bf 0611} (2006) 051.

\bibitem{Vera:2006un}
  A.~Sabio Vera, Nucl.\ Phys.\  B {\bf 746} (2006) 1.

\bibitem{Mueller:1986ey}
  A.~H.~Mueller and H.~Navelet, 
  Nucl.\ Phys.\  B {\bf 282} (1987) 727.

\bibitem{DelDuca:1993mn}
  V.~Del Duca and C.~R.~Schmidt, 
  Phys.\ Rev.\  D {\bf 49} (1994) 4510.

\bibitem{DelDuca:1994ng}
  V.~Del Duca and C.~R.~Schmidt, 
  Phys.\ Rev.\  D {\bf 51} (1995) 2150.

\bibitem{Stirling:1994zs}
  W.~J.~Stirling, 
  Nucl.\ Phys.\  B {\bf 423} (1994) 56.

\bibitem{Abachi:1996et}
  S.~Abachi {\it et al.}  [D0 Collaboration], 
  Phys.\ Rev.\ Lett.\  {\bf 77}, 595 (1996).

\bibitem{Abbott:1999ai}
  B.~Abbott {\it et al.}  [D0 Collaboration],
  Phys.\ Rev.\ Lett.\  {\bf 84}, 5722 (2000).

\bibitem{Abbott:1997nf}
  B.~Abbott {\it et al.}  [D0 Collaboration],
  Phys.\ Rev.\ Lett.\  {\bf 80}, 666 (1998).

\bibitem{Abe:1996mj}
  F.~Abe {\it et al.}  [CDF Collaboration],
  Phys.\ Rev.\ Lett.\  {\bf 77}, 5336 (1996)
  [Erratum-ibid.\  {\bf 78}, 4307 (1997)].

\bibitem{Lipatov:1985uk}
  L.~N.~Lipatov,
  Sov.\ Phys.\ JETP {\bf 63}, 904 (1986)
  [Zh.\ Eksp.\ Teor.\ Fiz.\  {\bf 90}, 1536 (1986)].

\bibitem{Bartels:2001ge}
  J.~Bartels, D.~Colferai and G.~P.~Vacca,
  Eur.\ Phys.\ J.\  C {\bf 24}, 83 (2002).

\bibitem{Bartels:2002yj}
  J.~Bartels, D.~Colferai and G.~P.~Vacca,
  Eur.\ Phys.\ J.\  C {\bf 29}, 235 (2003).

\bibitem{Kotikov:2000pm}
  A.~V.~Kotikov and L.~N.~Lipatov,
  Nucl.\ Phys.\  B {\bf 582}, 19 (2000).

\bibitem{Vera:2007kn}
  A.~Sabio Vera and F.~Schwennsen,
  Nucl.\ Phys.\  B {\bf 776} (2007) 170.

\bibitem{Marquet:2007xx}
  C.~Marquet and C.~Royon,
  arXiv:0704.3409 [hep-ph].

\bibitem{Vera:2007dr}
  A.~Sabio Vera and F.~Schwennsen,
  Phys.\ Rev.\  D {\bf 77} (2008) 014001.

\end{thebibliography}
\end{document}